\documentclass[3p,one column,fleqn]{elsarticle}
\biboptions{sort&compress}
\usepackage{amsmath,physics,amssymb,graphics}
\usepackage{xcolor}
\usepackage{hyperref}
\hypersetup{pdfauthor=author}
\makeatletter
\newcommand*{\rom}[1]{\expandafter\@slowromancap\romannumeral #1@}
\makeatother

\let\oldhat\hat

\renewcommand{\hat}[1]{\oldhat{\mathbf{#1}}}
\begin{document}
\begin{frontmatter}
	\title{A revision to the QES prescription}
	\author[mainaddress]{Amir A. Khodahami}
	\ead{a.khodahami@shirazu.ac.ir}
	\author[mainaddress]{Azizollah Azizi\corref{correspondingauthor}}
	\ead{azizi@shirazu.ac.ir}
	\address[mainaddress]{Department of Physics, Shiraz University, Shiraz 71949-84795, Iran}
	\cortext[correspondingauthor]{Corresponding author}
	\date{Received: date / Accepted: date}
	\begin{abstract}
		In this paper, we propose a revision to the Quantum Extremal Surface (QES) prescription, which plays a crucial role in describing the entanglement entropy of black holes. While derivations exist for the original QES prescription using the replica trick, they rely on a saddle point approximation that faces challenges when the number of replicas approaches unity. We highlight these challenges and develop a refined formula for the entanglement entropy that incorporates a weighted summation over multiple surfaces, contrasting the QES prescription's reliance on a single extremal surface. We validate our formulation by demonstrating consistency with expected results for both the early and late stages of black hole evaporation, confirming its alignment with the unitary evolution depicted by the Page curve.
	\end{abstract}
\end{frontmatter}
\raggedbottom
%%%%%%%%%%%%%%%%%%%%%%%%%%%%%%%%%%%%%%%%%%%%%%%%%%%%%%%%%%%%%%%%%%%%%%%%%%%%
%%%%%%%%%%%%%%%%%%%%%%%%%%%%%%%%%%%%%%%%%%%%%%%%%%%%%%%%%%%%%%%%%%%%%%%%%%%
\section{Introduction}
In 1976, Hawking demonstrated that the formation and evaporation processes of black holes lead to destruction of quantum information and hence violate the unitarity \cite{hawking1976a}\footnote{The information paradox can be understood as an inconsistency among unitarity, causality, and the semi-classical description of black hole evaporation.}. The extent of information loss is quantified by a concept known as ``fine-grained entropy'' or ``entanglement entropy'', defined as $S=-\Tr\rho\ln\rho$, where $\rho$ represents the density matrix of the black hole. In 2006, Ryu and Takayanagi introduced a holographic formula for the entanglement entropy \cite{ryu2006b,fursaev2006,nishioka2009b}. This formulation was further developed by \cite{hubeny2007b} to a covariant form and by \cite{faulkner2013b} to incorporate quantum corrections, eventually leading to the following formula proposed by \cite{engelhardt2015b}, which is applicable to evaporating black holes
\begin{equation}\label{Eq__EE_QES}
	S=\text{min}_X\left\{\text{ext}_X\left[\frac{\text{Area}(X)}{4G}+S_{\text{semi-cl}}\left(\Sigma_X\right)\right]\right\}.
\end{equation}
Here, $X$ is a codimension-2 surface and $\Sigma_X$ is the region bounded by $X$ and the surface from where we can almost assume that spacetime is flat, called cutoff surface\footnote{Determining the precise location of the cutoff surface is challenging according to the long-range nature of gravity \cite{almheiri2021b}.}. Moreover, the term $S_{\text{semi-cl}}(\Sigma_X)$ is standing for the entanglement entropy of the quantum fields on $\Sigma_X$ in the semi-classical description\footnote{The semi-classical description treats matter fields as quantum and the gravitational field as classical.}. The expression in the square brackets is called \textit{generalized entropy}, $S_\text{gen}(X)$. The surface that extremizes $S_\text{gen}(X)$ is called Quantum Extremal Surface (QES). According to the recipe by \cite{engelhardt2015b}, if there are multiple such quantum extremal surfaces, then we should select the one with the least $S_\text{gen}$.
\par
\begin{figure}[t]
	\centering
	\includegraphics[width=0.7\linewidth]{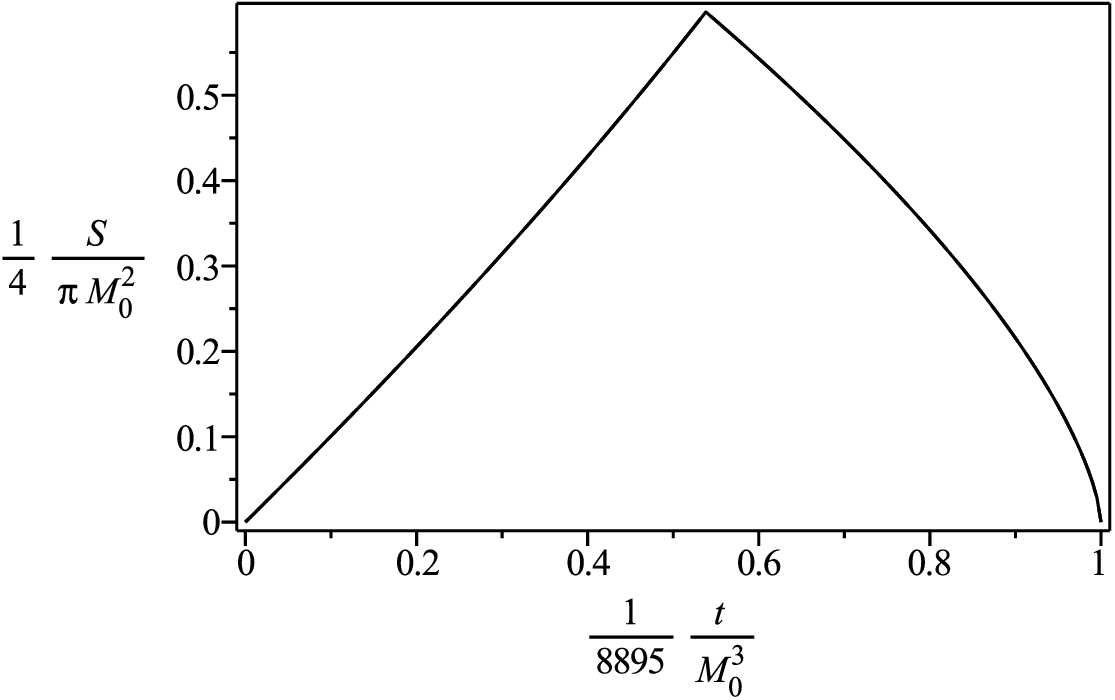}
	\caption{\label{f__Page} The Page curve, from \cite{page2013a}, depicts the entanglement entropy of an evaporating black hole initially in a pure state. This curve emerges from a comparison of the thermodynamic entropies associated with the radiation and the black hole over time. Specifically, it selects the entropy that is lower. The Page curve serves as a significant indicator of the unitarity of the black hole evaporation process.}
\end{figure}

 For an evaporating black hole that starts from a pure state, we identify two QESs as described by Eq.~\eqref{Eq__EE_QES}: (i) a vanishing surface with a growing contribution, and (ii) a nonvanishing surface just behind the event horizon with a decreasing contribution. In the early stages, the minimal QES is the vanishing one causing the generalized entropy to match the entanglement entropy of the bulk (inside the cutoff surface). Hence, the initial black hole's entanglement entropy starts at zero and continuously increases. On the other hand, the nonvanishing QES appears shortly after the black hole formation with a contribution that reflects the black hole's area, which decreases as the black hole evaporates. According to the minimal selection rule of Eq.~\eqref{Eq__EE_QES}, when the contribution of the nonvanishing surface becomes smaller than that of the vanishing one, it starts to represent the true entanglement entropy of the black hole \cite{almheiri2021b}. This switch between the two QESs allows the black hole's entanglement entropy to closely follow the Page curve shown in Fig.~\ref{f__Page}, which is indicative of unitary evaporation \cite{page1983b,page1993a,page2013a,almheiri2019,penington2020,Marolf2020,Dong2020,Almheiri2020a,Kudler-Flam2022} (see \cite{harlow2016} for further discussions).
\par R\'enyi entropy is defined as
\begin{equation}
	S_n:=\frac{1}{1-n}\ln\left(\frac{\Tr(\rho^n)}{\left(\Tr\rho\right)^n}\right),
\end{equation}
for an unnormalized density matrix $\rho$. In the limit $n\to1$, it reduces to the entanglement entropy $S=-\Tr\rho\ln\rho$, i.e.,
\begin{equation}\label{Eq__Rep}
	S = \lim_{n \to 1} \frac{1}{1-n} \ln\left(\frac{\tr(\rho^n)}{(\tr\rho)^n}\right).
\end{equation}
The replica trick is a method used to compute entanglement entropy by exploiting the properties of R\'enyi entropy. The trick involves three steps: (i) compute $S_n$ for an integer $n>1$, (ii) analytically continue $n$ to non-integer values, and (iii) take the limit $n\to1$ to obtain the entanglement entropy. This approach circumvents the need to directly compute the nonlinear logarithm, offering a simpler way to calculate the entanglement entropy \cite{dong2016}. However, there are some costs associated with this simplification, which we will now discuss: (1) There is no guarantee that the R\'enyi entropy is always analytic. (2) There is no guarantee that the analytic form for non-integer $n$ is unique\footnote{In ordinary quantum mechanics, $\Tr \rho^n$ admits a unique analytic continuation. In quantum field theory, however, the density matrix formalism is not rigorously defined, so the existence and uniqueness of such a continuation are not guaranteed \cite{witten2025}.}. There might be some terms like $(1/(1-n)!)$ that are zero for every integer $n>1$ but contribute to the limit $S=\lim_{n\to1}S_n$. These terms can complicate the analytic continuation and introduce ambiguities in the process \cite{headrick2010,witten2025}. (3) Even though the R\'enyi entropy gives the entanglement entropy in the limit $n\to1$, it has an infinite derivative at this point and is not differentiable
\begin{equation}\label{Eq__Infinite_dR}
	\left.\frac{d}{dn}S_n\right|_{n\to1}=\left.\frac{S}{2(n-1)}\right|_{n\to1}\to\infty.
\end{equation}
Hence, although the limit exists, the R\'enyi entropy is not expandable in the neighborhood of $n=1$.
\par The authors in \cite{almheiri2020b} and \cite{penington2022b} investigated the QES prescription within a JT-gravity framework, as JT-gravity acts as a two-dimensional toy model for quantum gravity (see \cite{mertens2023} for a review). Using the replica trick enables one to consider all contributions to the entanglement entropy by summing over different topologies. In the two-replica system, two distinct topologies arise: one disconnected, which corresponds to the empty QES, and the other a non-trivial connected topology, which corresponds to the non-vanishing QES. The authors subsequently revisited the QES prescription using a saddle point approximation\footnote{The information paradox and its resolution is also worked out in de Sitter and asymptotically flat spacetimes \cite{Marolf2021,Marolf2021a,Geng2021,Arefeva2021,Hashimoto2020,Wang2021,Gautason2020,hartman2020a}.}. However, subtleties arise when sending the number of replicas $n$ to 1, as specified by the replica formula. Addressing these subtleties alters the QES prescription through invalidating the saddle-point approximation used in its derivation.
\par This paper is structured as follows: In Sec.~\ref{sec-review}, we present a concise overview of the derivation of the QES prescription as outlined in \cite{almheiri2020b}. In Sec.~\ref{sec-challenges}, we explore the challenges that emerge when approaching the limit as $n\to1$. To address these challenges, we propose a more refined treatment in Sec.~\ref{Sec__Revision}, that leads to a revised formula for the entanglement entropy. This new formula utilizes a weighted summation over multiple surfaces, in contrast to the QES prescription’s reliance on just the extremal surface. Under certain approximations, the revised formula converges to the generalized entropy at an extremal surface that extremizes the semi-classical entropy. In Sec.~\ref{sec-info}, we provide an information-theoretic interpretation of the weighting and clarify the physical meaning of the extremal surface. In Sec.~\ref{sec-check}, we investigate the revised formula, validating its consistency with unitarity. Finally conclusions are presented in Sec.~\ref{sec-conc}.
\section{A review of the derivation of the QES prescription}\label{sec-review}
In \cite{almheiri2020b}, the authors consider a model of an $AdS_2$ black hole glued to flat space. This model is constructed by an $AdS_2$ JT-gravity theory coupled to a two-dimensional CFT. The CFT also lives in non-gravitational Minkowski regions, and has transparent boundary conditions at the $AdS$ boundary. The dilaton goes to infinity at the $AdS_2$ boundary so it is consistent to freeze gravity on the outside. The Euclidean action for this theory reads as
\begin{equation}
-I=\frac{\phi_0}{16 \pi G}\left[\int_{\Sigma_2} R+\int_{\partial \Sigma_2} 2K\right] + \int_{\Sigma_2} \frac{\phi}{16 \pi G}(R+2)+\frac{\phi_b}{16 \pi G} \int_{\partial \Sigma_2} 2 K+I_{\text{CFT}}[g].
\end{equation}
Here, $\phi_0/16\pi G$ ($=S_0/4\pi$) is a dimensionless normalization constant that determines the scale of the gravitational action. The term $\Sigma_2$ denotes the $AdS_2$ manifold, including its boundary $\partial\Sigma_2$. The Ricci scalar $R$ corresponds to the metric tensor $g_{\mu\nu}$, while $K$ represents the extrinsic curvature of the boundary. The dilaton field is denoted by $\phi$, with $\phi_b$ indicating its value at the boundary. Lastly, $I_{\text{CFT}}[g]$ represents the action of the CFT. Note that while \cite{almheiri2020b} sets $4G=1$, we retain explicit factors of $G$ for some order of magnitude analysis purposes.
\begin{figure*}[t]
	\centering
	\includegraphics[width=0.85\textwidth]{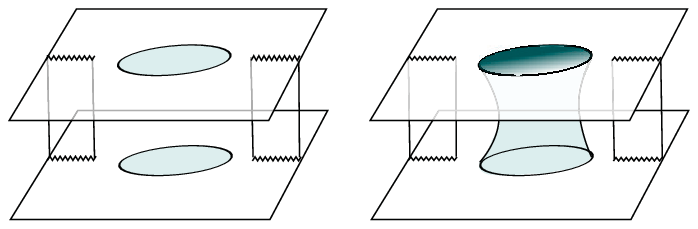}
	\caption{\label{f__R2} Illustrating diagrams are given in \cite{almheiri2020b} for two different contributions of $\Tr\rho^2$. Left is what Hawking considered in his calculations and hence is called Hawking saddle. It simply considers the black holes as evaporating independently and is the dominant contribution before the Page time. Right is a non-trivial topology in which the interiors of the black holes are connected forming a wormhole called replica wormhole. It becomes the dominant contribution after the Page time.}
\end{figure*}
\par In the replicated system, the geometry of the gravitational region is not predetermined. Rather, it is dynamically established by fixing the boundary conditions, which allows gravity to fill in the geometry. The authors obtained two distinct topologies for the two-replica system, as illustrated in Fig.~\ref{f__R2}. The action for the replicated system is expressed as\footnote{In \cite{almheiri2020b}, the authors explored scenarios involving multiple $w$'s, corresponding to gluing surfaces composed of two or more segments. To ensure clarity and conciseness, we have opted to omit this summation while still preserving the generality of our arguments.}
\begin{equation}\label{Eq__In}
	\begin{split}
		-I_{n}= n\,\frac{\phi_0}{16 \pi G}\left[\int_{\Sigma_2} R+\int_{\partial \Sigma_2} 2 K\right]+&n\int_{\Sigma_2} \frac{\phi}{16 \pi G}\left(R+2\right)\\+&n\,\frac{\phi_b}{16 \pi G} \int_{\partial \Sigma_2} 2 K-\left(n-1\right)\left[\frac{\phi_0+\phi\left(w\right)}{4G}\right] +n\,I_{\text{CFT}}[g],
	\end{split}
\end{equation}
with $w$ denoting the location of the so-called twist operator, corresponding to a gluing surface in the replicated geometry\footnote{The twist operator is inserted at the fixed point of the replica symmetry in the quotient manifold $\mathcal{M}_n/\mathbb{Z}_n$. It corresponds to the gluing surface in the original replicated manifold $\mathcal{M}_n$.}. Following the results of \cite{almheiri2020b,penington2022b}, the action for the replicated system near $n=1$ simplifies to
\begin{equation}\label{Eq__InIS}
	I_n=nI+\left(n-1\right)S_\text{gen}(w),
\end{equation}
which is the full off-shell action that we need to extremize to find the classical solution. In this way, we obtain the QES prescription using Eq.~\eqref{Eq__Rep}.
\section{Challenges in the $n\to1$ limit}\label{sec-challenges}
The derivation reviewed in the previous section relies on a saddle point approximation, which fails to hold in the desired limit as $n \to 1$. In this context, the variable $w$, introduced in the replicated action $I_n$, is assumed to be fixed according to the classical equation that extremizes the generalized entropy
\begin{equation}
	(n-1)\frac{\partial}{\partial w}S_\text{gen}(w)=0.
\end{equation}
Notably, this equation is of $\mathcal{O}(n-1)$, and therefore becomes automatically satisfied as we approach the limit $n \to 1$. This occurs because, in Eq.~\eqref{Eq__InIS}, the generalized entropy is accompanied by a prefactor of $(n-1)$. Consequently, the path integral receives equal contributions from all twist operator locations due to the vanishing $(n-1)$ prefactor.
\par One might argue that the same saddle-point approximation is used in deriving the Ryu-Takayanagi (RT) formula. However, the RT formula represents the $G\to0$ limit of the QES prescription, where the saddle-point approximation is justified. This is because the generalized entropy contains an $\text{Area}/4G$ term, which sharply localizes the path integral around the minimal surface when $G\to0$. To clarify, one can first take $G\to0$ and apply the saddle-point approximation, and then substitute the result into the replica formula to derive the RT formula \cite{lewkowycz2013b,witten2025}. In contrast, for the QES prescription, $G$ is kept small but finite. Here, the parameter $G/(n-1)$ governs the width of the integrand. As $n\to1$, this width diverges for finite $G$, meaning all possible locations of the twist operator contribute equally, and no single configuration is preferred.
\par One way to justify the derivation in the previous section is to introduce a cutoff $\epsilon$ for $(n-1)$ by replacing $n \to 1$ with $n \to (1 + \epsilon)$ in the replica formula Eq.~\eqref{Eq__Rep}. Since $G$ is extremely small ($\mathcal{O}(10^{-70})$ in squared meters), there exists a sufficiently small $\epsilon$ such that
\begin{equation}
	\epsilon \frac{\text{Area}}{G} \gg 1,
\end{equation}
for a typical area scale (e.g., the horizon area of a solar-mass black hole is $\mathcal{O}(10^7)$ squared meters). Under this condition, the saddle-point approximation remains valid because the integrand becomes sharply peaked (i.e., the width $G/\epsilon$ is very small). While this condition ensures the validity of the saddle-point approximation, the divergent behavior of the R\'enyi entropy derivative at $n=1$ (Eq.~\eqref{Eq__Infinite_dR}), introduces errors comparable in magnitude to the entanglement entropy itself when a cutoff is introduced. Therefore, the limit $n \to 1$ must be taken with infinite precision, rendering the use of any cutoff invalid.
\par Notably, general arguments based solely on the action fail to yield a valid entanglement entropy formula, even without invoking the saddle-point approximation. To demonstrate this explicitly, we bypass the approximation and consider all possible values of $w$. Using Eq.~\eqref{Eq__InIS}, we obtain
\begin{equation}
	\frac{\tr(\rho^n)}{\left(\tr\rho\right)^n} \approx \sum_w \exp\left(-(n-1)S_\text{gen}(w)\right).
\end{equation}
Substituting this into the replica formula (Eq.~\eqref{Eq__Rep}) gives
\begin{equation}
	S = \sum_w S_\text{gen}(w),
\end{equation}
which is clearly incorrect. A more careful approach, which takes into account the distinct integration domains associated with each term in the replicated action, is therefore required to obtain the appropriate expression for the entropy. This will be addressed in the following section.
\section{Revising the entanglement entropy formula}\label{Sec__Revision}
To identify a suitable formula for the desired limit $n\to1$, it is crucial to note that the classical equations arising from the replicated action $I_n$ in Eq.~\eqref{Eq__In} are mainly of $\mathcal{O}(n)$, except for the equation governing $w$, which is of $\mathcal{O}(n-1)$ \cite{almheiri2020b}. Consequently, the saddle point approximation can be applied to all fields except the variable $w$. For a given choice of gluing surface, the degrees of freedom in the entire geometry can be divided into two parts: (i) the bulk that lies within the gluing surface, and (ii) the regions that extend outside the gluing surface. Eq.~\eqref{Eq__InIS} indicates that the degrees of freedom contained within the gluing surface are characterized by $(n-1)$ times the generalized entropy, while those outside are described as $n$ times the original action $I$. The coefficients of $(n-1)$ and $n$ stemming from the analytic continuation $(2 \to n)$ also reflect this distinction. Specifically, the degrees of freedom in (i) arise from selecting $2$ out of $n$ replicas (which is proportional to $(n-1)$), multiplied by the generalized entropy. In contrast, those in (ii) originate from each replica contributing its own action, hence being proportional to $n$. Considering this distinction in writing the path integral for $n \sim 1$, we have
\begin{equation}
			\tr(\rho^n)=\sum_w \left(\int_{\text{region (i)}}+\int_{\text{region (ii)}}\right)\exp(-I_n(w))
			=\sum_w\exp(-(n-1)S_\text{gen}(w))\int_{\text{region (ii)}}\hspace{-6 mm}\exp(-nI).
\end{equation}
The integral over region (ii) can be approximated by its on-shell value within the domain enclosed by $w$ and the cutoff surface. Denoting this region by $\Sigma_w$, we obtain
	\begin{equation}\label{eq:pathint}
		\tr(\rho^n)=\sum_w\exp(-(n-1)S_\text{gen}(w))\exp(-n\widetilde{I}(\Sigma_{w})),
	\end{equation}
where $\widetilde{I}(\Sigma_w)$ is the Euclidean on-shell action evaluated on $\Sigma_w$. This action can be written in terms of a thermodynamic potential $\Omega(\Sigma_w)$ as
\begin{equation}
	\widetilde{I}(\Sigma_{w})=\beta\Omega(\Sigma_{w}),
\end{equation}
where $\beta = 1/T$ denotes the inverse temperature. To identify the relevant potential $\Omega(\Sigma_w)$, one must make explicit what is held fixed in the path integral. The choice of the gluing surface $w$ determines how the spacetime is partitioned and thereby fixes the entanglement data associated with the cut. In particular, the entanglement entropy across $w$ is fixed by the boundary conditions imposed at the gluing surface, rather than being treated as an independent integration variable.\footnote{The same logic applies at the cutoff surface: the state of the region beyond the cutoff fixes the entanglement between $\Sigma_w$ and the exterior.} Accordingly, the path integral over $\Sigma_w$ sums over all bulk matter-field and metric configurations consistent with these boundary conditions, i.e.\ at fixed entanglement data, while all other local degrees of freedom fluctuate. Decomposing the thermodynamic entropy into a classical component and an entanglement part $S=S_{\text{cl}}+S_{\text{ent}}$, the thermodynamic potential takes the form
\begin{equation}
	\Omega(\Sigma_w) = E-TS_{\text{cl}}-\cdots=TS_{\text{ent}}(\Sigma_w) ,
\end{equation}
where the ellipsis accounts for additional contributions---such as mechanical terms arising from changes in the volume induced by variations of the metric, and chemical-potential terms associated with matter-field configurations carrying different particle numbers---which are inherently encoded in the thermodynamic potential. Substituting this result back into Eq.~\eqref{eq:pathint} yields
\begin{equation}\label{Eq__pathint}
	\tr(\rho^n) = \sum_w \exp\left(-(n-1)S_\text{gen}(w)\right)\exp\left(-nS_{\text{ent}}(\Sigma_w)\right).
\end{equation}
Now we can use the replica formula (Eq.~\eqref{Eq__Rep}) to find the entanglement entropy of the radiation
\begin{equation}\label{Eq__S}
		S=\frac{\sum_w\left[S_\text{gen}(w)+S_{\text{ent}}(\Sigma_w)\right]\exp(-S_{\text{ent}}(\Sigma_w))}{\sum_w\exp(-S_{\text{ent}}(\Sigma_w))}+\ln(\sum_w\exp(-S_{\text{ent}}(\Sigma_w))).
\end{equation}
The first term can be interpreted as the average of the quantity inside the brackets, with a probability distribution proportional to $\exp(-S_{\text{ent}}(\Sigma_w))$
\begin{equation}\label{Eq__S_2}
	S=\langle S_\text{gen}(w)\rangle_w+\langle S_{\text{ent}}(\Sigma_w)\rangle_w+\ln(\sum_w\exp(-S_{\text{ent}}(\Sigma_w))).
\end{equation}
This probability distribution is not immediately evident from the form of the replicated action $I_n$; rather, it arises from the limits of the path integral over region (ii), and not only from the integrand itself. Consequently, a more detailed analysis was required to demonstrate this, going beyond general arguments based solely on the action, as emphasized in \cite{almheiri2020b} and \cite{penington2022b}. This weight plays a crucial role in the convergence of the obtained formula, as will be explored in Sec.~\ref{sec-check}.
\par The expression for the entanglement entropy in
Eq.~\eqref{Eq__S_2} can be significantly simplified under certain approximations. In a UV-complete theory of quantum gravity,
the entropy associated with the region $\Sigma_w$ generally receives
contributions from both fundamental quantum-gravitational microstates,
such as Planckian degrees of freedom, and low-energy effective fields.
Within the framework of gravitational Effective Field Theory (EFT), the
UV-sensitive gravitational contributions can be absorbed into the
renormalization of the gravitational couplings, while the remaining
non-perturbative quantum-gravity corrections are suppressed at leading
semi-classical order. Therefore, the formula can be generalized to broader cases, including the four-dimensional Schwarzschild black hole of significant interest, by replacing $S_{\text{ent}}(\Sigma_w)$ with $S_\text{semi-cl}(\Sigma_X)$, where the physical domain of the gluing surface is naturally bounded by the singularity and the event horizon, $X \in [0, r_{\text{Sch}}]$, yielding
\begin{equation}\label{Eq__S__3}
	S=\frac{\sum_{X\leq r_\text{Sch}}\left[S_\text{gen}(X)+S_\text{semi-cl}(\Sigma_X)\right]\exp(-S_\text{semi-cl}(\Sigma_X))}{\sum_{X\leq r_\text{Sch}}\exp(-S_\text{semi-cl}(\Sigma_X))}+\ln(\sum_{X\leq r_\text{Sch}}\exp(-S_\text{semi-cl}(\Sigma_X))).
\end{equation}
Within this approximation, the probability distribution, $P(X)\propto \exp(-S_{\text{semi-cl}}(\Sigma_X))$, assigns a thermal-like statistical weight to the possible locations of the gluing surface. In the thermodynamic limit, where
fluctuations in the position of the gluing surface are suppressed, this
distribution is expected to be sharply peaked around a dominant
configuration $X=\widetilde{X}$, which minimizes $S_{\text{semi-cl}}(\Sigma_X)$. Within this effective description, the backreaction of quantum matter fields, encoded in their semi-classical entanglement entropy, plays the
primary role in stabilizing the position of the gluing surface. In this
case, the last two contributions in
Eq.~\eqref{Eq__S__3} cancel each other at leading order, and the entanglement entropy reduces to
\begin{equation}\label{Eq__Sapp}
	S\approx \langle S_\text{gen}(X) \rangle_X \approx S_\text{gen}(\widetilde{X}).
\end{equation}
Thus, within the single-saddle approximation, the entanglement entropy is given by the generalized entropy evaluated on the dominant configuration. This result is QES-like, although the relevant surface is selected here by
the minimization of the semi-classical matter contribution rather than
by directly extremizing the complete generalized entropy.
\par This physical picture also clarifies the regime in which the localized, single-saddle description must break down. Near the Page transition, the Hawking-like and replica-wormhole-like configurations
can become nearly degenerate, so that no single surface dominates the
statistical sum. In this crossover regime, the location of the gluing
surface is intrinsically delocalized, and the full expression in
Eq.~\eqref{Eq__S__3} must be retained rather than its single-saddle
approximation. The entropy is then determined by the collective
competition of the relevant saddles, which provides a smooth
interpolation between the two asymptotic phases and replaces the abrupt
switch of the strict $\min\operatorname{ext}$ prescription by a
continuously regulated Page-time crossover.
\par Before closing this section, we would like to express some points regarding the revised entanglement entropy formula, Eq.~\eqref{Eq__S__3}. First, it is clear from its derivation that the inclusion of higher replicas alters the expression within the square brackets while leaving the weight $\exp(-S_\text{semi-cl}(\Sigma_X))$ unchanged. This occurs because these higher replicas would contribute to the path integral with coefficients that vanish as sending the number of replicas $n$ to unity. Hence the location of the extremal surface is almost unchanged while the value of the entanglement entropy itself changes.
\par We also emphasize that the gluing surface $w$ is, in general, a continuous variable. Accordingly, the symbol $\sum_w$ should be understood as a shorthand for an integral over the space of admissible gluing surfaces,
\begin{equation}
	\sum_w \;\equiv\; \int \mathcal{D}w\,\mu[w] \, ,
\end{equation}
where $\mu[w]$ denotes the corresponding measure. In the semi-classical regime, this functional integral is evaluated by a saddle-point approximation and therefore effectively reduces to a discrete sum over the dominant saddle surfaces. The leading contribution is controlled by the exponential weight in Eq.~\eqref{Eq__S__3}, while the measure $\mu[w]$ contributes only through subleading prefactors. In particular, the convergence of the integral is governed primarily by the behavior of the effective exponent, whereas the measure affects the overall normalization. Hence, as long as the effective exponent is bounded from below---which is the case in Euclidean signature---the integral is well-defined and the leading result is not sensitive to the detailed choice of $\mu[w]$. Quantitative measure dependence may become relevant only near crossover regions, such as around the Page transition, where competing saddles carry comparable weights (see also \cite{Dong2020,Akers2021}).
\par Finally, we formulate the present analysis in Euclidean signature. The basic reason is that the Euclidean path integral carries a damping weight, so that dominant configurations can be identified through a standard saddle-point approximation while subleading contributions are comparatively well controlled. This provides a firmer foundation for the statistical interpretation of our weighting procedure: the candidate surfaces may be viewed as forming an effective ensemble, from which a dominant contribution emerges, yielding a QES-like description when the distribution becomes sharply peaked. By contrast, a direct Lorentzian formulation is considerably more subtle (see \cite{blommaert2023} for a detailed discussion). Lorentzian path integrals are oscillatory rather than damping, making convergence less transparent and the associated semiclassical interpretation more delicate. In practice, one often relies on Euclidean methods followed by analytic continuation, or on a more careful contour deformation in complexified field space. For these reasons, the revised formula here should be regarded primarily as a Euclidean semiclassical diagnostic, expected to be reliable in regimes where the distribution is localized and fluctuations around the leading configuration are parametrically suppressed. In this sense, the Euclidean framework is not merely a technical convenience, but an appropriate approximation for isolating the dominant entropic structure while postponing the substantially more intricate problem of a fully Lorentzian formulation.
\section{Information-theoretic interpretation of the weighting}\label{sec-info}
The probability distribution appearing in our refined entropy formula admits a natural interpretation in terms of information theory and spacetime connectivity. This interpretation is useful because it makes explicit what is being optimized in the weighted sum over candidate surfaces. Rather than selecting a single surface by imposing an extremization condition from the outset, the present framework assigns a statistical weight to each candidate configuration and allows the dominant contribution to emerge dynamically from the competition among them.

To make this structure more transparent, we multiply the numerator and denominator of the normalized probability distribution by $\exp(S(\Sigma_0))$, which gives
\begin{equation}
	P(X)=\frac{\exp(S(\Sigma_0)-S(\Sigma_X))}{\sum_X\exp(S(\Sigma_0)-S(\Sigma_X))}.
\end{equation}
Here $\Sigma_0$ denotes the reference configuration with no nontrivial bulk excision, while $\Sigma_X$ labels the candidate surface associated with the parameter $X$. The quantity $S(\Sigma_0)-S(\Sigma_X)$ therefore measures how much entropy is effectively removed, reorganized, or reattributed when one passes from the reference configuration to the configuration associated with $X$. Thus, the weighting should be regarded as a comparison between different decompositions of the same reference configuration, not as an independent probability assigned to an isolated geometric surface. This expression can be rewritten in a more suggestive way by identifying the bulk region enclosed by the surface $X$ as $\mathcal{B}_X$. By construction, one may decompose the reference region as
\begin{equation}
	\Sigma_0=\mathcal{B}_X\cup\Sigma_X,
\end{equation}
for any admissible choice of $X$. With this decomposition, the entropy difference appearing in the exponent can be written as
\begin{equation}
	S(\Sigma_0)-S(\Sigma_X)
	=
	S(\mathcal{B}_X\cup\Sigma_X)-S(\Sigma_X),
\end{equation}
which is precisely the quantum conditional entropy of the candidate bulk region with respect to its complement, $S(\mathcal{B}_X|\Sigma_X)$. The appearance of a conditional entropy follows directly from the fact that each candidate surface defines a bipartition of the reference configuration. This provides a conditional version of the usual entropic selection principle. In ordinary statistical mechanics, a macrostate with entropy $S$ carries a statistical weight proportional to $\exp(S)$, reflecting the number of microscopic configurations compatible with it. In the present case the candidate bulk region $\mathcal{B}_X$ is not an isolated system; it is correlated with the complementary sector $\Sigma_X$. Therefore the relevant degeneracy is not the unconstrained degeneracy $\exp[S(\mathcal{B}_X)]$, but the effective conditional degeneracy
\begin{equation}
	\Omega(\mathcal{B}_X|\Sigma_X)
	=
	\exp\!\left[S(\mathcal{B}_X|\Sigma_X)\right].
\end{equation}
Equivalently, using the definition of mutual information,
\begin{equation}
	I(\mathcal{B}_X;\Sigma_X)=S(\mathcal{B}_X)+S(\Sigma_X)-S(\mathcal{B}_X\cup\Sigma_X),
\end{equation}
the probability distribution becomes
\begin{equation}
		P(X)=\frac{\exp(S(\mathcal{B}_X)-I(\mathcal{B}_X;\Sigma_X))}{\sum_X\exp(S(\mathcal{B}_X)-I(\mathcal{B}_X;\Sigma_X))}\propto \exp(S(\mathcal{B}_X)-I(\mathcal{B}_X;\Sigma_X)).
\end{equation}
The first contribution, $S(\mathcal{B}_X)$, measures the amount of information loss associated with the bulk region enclosed by the surface. Equivalently, it is the logarithm of the number of microscopic configurations compatible with that region; hence, larger $S(\mathcal{B}_X)$ enhances the statistical weight of surfaces that capture more bulk degrees of freedom. The second contribution, $I(\mathcal{B}_X;\Sigma_X)$, plays the opposite role. Mutual information measures the total amount of correlation shared between the enclosed bulk region, $\mathcal{B}_X$, and the complementary, $\Sigma_X$. It is therefore natural to interpret this term as a correlation cost. A surface that cuts through strongly correlated degrees of freedom is penalized, because the information across that interface is not cleanly separated into independent subsystems. In other words, even if a candidate surface encloses many microstates, it does not receive a large weight if doing so requires cutting across substantial correlations between the enclosed region and its complement.

\par The resulting competition has a clear physical meaning: the preferred surface is not simply the one that encloses the largest entropy, nor simply the one that minimizes correlations, but rather the one that optimally balances these two tendencies. This provides an information-theoretic characterization of the dominant surface selected by the weighted sum. In the regime where the probability distribution is sharply peaked, this dominant surface coincides with the effective extremal surface $\widetilde{X}$ appearing in Eq.~\eqref{Eq__Sapp}, thereby recovering a QES-like description as an emergent saddle of the probability distribution rather than as a variational condition imposed a priori.

\par From this perspective, our interpretation is complementary to previous information-theoretic viewpoints on islands and entanglement-wedge reconstruction. For the dominant surface, the enclosed bulk region $\mathcal{B}_X$ is naturally interpreted as the island region. From this viewpoint, the island is selected because it captures a large amount of entropy while remaining only weakly correlated with the black hole interior across the cut. This is consistent with the idea that the island should be regarded as part of the entanglement wedge of the radiation rather than of the naively defined black hole interior \cite{engelhardt2023,engelhardt2022,maldacena2013,antonini2025}. Previous information-theoretic approaches emphasize the encoding and reconstructability of bulk information in the radiation or boundary degrees of freedom. In the quantum-error-correction interpretation, this reconstructability is an algebraic statement about representing bulk operators on the radiation or boundary algebra within an appropriate code subspace. The present formulation addresses a related but distinct question: given a family of admissible semiclassical cuts, which bipartition is statistically preferred? The mutual-information contribution connects these viewpoints by making the correlations across each candidate cut explicit. Thus, the mutual-information term does not by itself determine the entanglement wedge, but it incorporates the correlation structure relevant to an island-like assignment into the relative weighting of candidate surfaces. This interpretation is also closely related to the broader connection between spacetime connectivity and quantum information. As emphasized in \cite{raamsdonk2010a,raamsdonk2010}, correlations encoded by mutual information provide a diagnostic of how spacetime regions are effectively glued together in the semiclassical description. Although our setting is not identical to the original holographic constructions, the same qualitative lesson applies: mutual information provides a useful diagnostic of the strength of correlations, and changes in these correlations can be interpreted semiclassically as changes in the effective connectivity between the corresponding regions. In the present context, the suppression factor $\exp[-I(\mathcal{B}_X;\Sigma_X)]$ therefore favors configurations in which the enclosed region is only weakly tied to the black hole interior across the candidate cut, precisely as expected when an island region is reorganized as part of the radiation. The proposed information-theoretic weighting is also naturally aligned with recent discussions on the Page curve, the monogamous redistribution of quantum entanglement in curved spacetimes \cite{wu2022, wu2023}, and the conservation of mutual information during black hole evaporation \cite{wu2024}. In our framework, the competition between candidate surfaces is regulated by the mutual information, which penalizes cuts that separate strongly correlated degrees of freedom. This provides a semiclassical perspective on how the redistribution of correlations dictates the dominance of different island-like regions. In this sense, the formalism offers additional physical insight into our results: the switch of the dominant surface can be viewed as an information-theoretic transition in which correlations are reorganized rather than lost, in close analogy with the Page-curve picture and the monogamous sharing of entanglement.

\par It is important, however, to state the scope of this interpretation carefully. The quantity $P(X)$ should not be viewed as a fundamental microscopic probability distribution derived from a complete theory of quantum gravity; rather, within our semiclassical Euclidean framework, it is an effective probability weight that parametrizes the relative importance of different candidate gluing surfaces. Accordingly, the information-theoretic interpretation proposed here should be understood as a semiclassical diagnostic of the competition among saddles. Within that regime, it provides a useful and physically intuitive way to understand why the dominant contribution is associated with an island-like region and why the refined entropy formula naturally goes beyond a strictly single-surface prescription. This interpretation also makes clear that our proposal differs conceptually from the standard QES prescription. In the latter, one directly extremizes the generalized entropy over candidate surfaces. In contrast, here the dominant surface emerges from a weighted competition among configurations. The extremal description is recovered only when the distribution becomes sufficiently sharp.

\section{Validating the new formula}\label{sec-check}
Let check whether the obtained formula for the entanglement entropy, Eq.~\eqref{Eq__S__3}, or simply its approximate form Eq.~\eqref{Eq__Sapp}, gives the intended results. We assume that the evaporation process has started at $t_\text{start}$, reached its half at $t_\text{half}$---analogous to the Page time---and completed at $t_\text{evap}$. For $t\ll t_\text{half}$, the formula should give the Hawking saddle, while for $t\gg t_\text{half}$, it should reflect the replica wormhole. Excluding the very initial times according to the last paragraph of Sec.~\ref{Sec__Revision}, a macroscopic number ($N>10^{23}$) of Hawking particles are present within the black hole interior when $t \ll t_\text{half}$, with their pairs located within the cutoff surface. This results in a rapid increase in $S_\text{semi-cl}(\Sigma_X)$ as a function of $X$, i.e.,
\begin{equation}
	S_\text{semi-cl}(\Sigma_X)\approx \mathcal{O}\left(\frac{NX}{r_\text{Sch}}\right).
\end{equation}
Consequently, $\exp(-S_\text{semi-cl}(\Sigma_X))$ decreases at a very fast rate, resulting in substantial contributions to the entanglement entropy only from surfaces that are nearly vanishing ($X \approx 0$). So we obtain
\begin{equation}
		S(t \ll t_\text{half}) \approx S_\text{gen}(X=0) = S_\text{semi-cl}(\Sigma_{X=0}),
\end{equation}
which is the expected Hawking saddle. At the other limit of interest, i.e., when $t\gg t_\text{half}$, a significant number of Hawking particles have already left the cutoff surface. Consequently, $S_\text{semi-cl}(\Sigma_X)$ is very large for any choice of $X$ other than the Schwarzschild radius of the black hole, $r_\text{Sch}$. Only when $X\approx r_\text{Sch}$ we see a non-negligible contribution from $\exp(-S_\text{semi-cl}(\Sigma_X))$, as this choice excludes pairs of those particles that have left the cutoff surface. Therefore, $\exp(-S_\text{semi-cl}(\Sigma_X))$ is vanishingly small for any $X\neq r_\text{Sch}$ and is of $\mathcal{O}(1)$ for $X\approx r_\text{Sch}$. Hence Eq.~\eqref{Eq__Sapp} gives
\begin{equation}
		S(t\gg t_\text{half}) \approx S_\text{gen}(X=r_\text{Sch})\approx \frac{\text{Area}(X=r_\text{Sch})}{4G},
\end{equation}
which is the desired replica wormhole.
\par An especially nontrivial regime is the neighborhood of the Page
time, $t\sim t_{\text{half}}$, where the single-saddle approximation
underlying Eq.~\eqref{Eq__Sapp} ceases to be parametrically reliable.
Indeed, as the evaporation proceeds, the Hawking-like configuration
localized near $X\approx 0$ and the replica-wormhole-like configuration
localized near $X\approx r_{\mathrm{Sch}}$ may become nearly degenerate
at the level of the semi-classical matter contribution, so that
$S_{\mathrm{semi-cl}}(\Sigma_{X\approx 0})$ and
$S_{\mathrm{semi-cl}}(\Sigma_{X\approx r_{\mathrm{Sch}}})$ differ only
subleadingly. In such a situation, the distribution
$P(X)\propto \exp(-S_{\mathrm{semi-cl}}(\Sigma_X))$ loses its sharp
localization and is expected to develop a bimodal profile, indicating
the simultaneous relevance of two competing gluing surfaces. The
entropy can then no longer be approximated by the generalized entropy
evaluated on a single dominant configuration; rather, one must retain
the full expression in Eq.~\eqref{Eq__S__3}, which effectively performs
a statistical summation over the candidate saddles. This provides a
natural interpolation between the early-time Hawking regime and the
late-time replica wormhole regime, and in particular suggests that the
proposed framework replaces the abrupt saddle exchange of the standard
$\min\operatorname{ext}$ prescription by a smoother Page-time crossover
controlled by the relative statistical weights of the competing
semi-classical configurations.
\section{Conclusions}\label{sec-conc}
The QES prescription is a well-established framework crucial for describing the entanglement entropy of black holes. However, derivations employing the replica trick rely on a saddle point approximation, which encounters difficulties as the number of replicas approaches unity. We have highlighted these challenges and developed a refined formula for the entanglement entropy. This revised formula incorporates a weighted summation over multiple surfaces, contrasting with the QES prescription, which depends on a single extremal surface. We validated our formulation by demonstrating its consistency with unitarity.

The statistical weight has a direct information-theoretic interpretation. It favors configurations that contain a large number of accessible microstates while suppressing surfaces that separate strongly correlated degrees of freedom. Equivalently, the mutual-information contribution acts as a soft correlation constraint on the allowed bipartitions. The dominant configuration is therefore selected by the balance between its statistical multiplicity and the correlations cut by the corresponding surface. This provides an additional physical interpretation of the transition between the Hawking-like and replica-wormhole-like regimes. When the weights of two candidate configurations become comparable, the relevant correlations are redistributed between competing semiclassical partitions. The associated change in the dominant contribution can thus be viewed as an information-theoretic transition, rather than as the loss of information. In this sense, the weighted formulation gives a smoother description of the Page-time crossover than the abrupt exchange implied by a strict $\min\operatorname{ext}$ prescription.

Although the present analysis is formulated for an evaporating Schwarzschild black hole, the underlying weighted construction is not conceptually restricted to this particular geometry. Its basic ingredients are the statistical multiplicities of candidate gluing surfaces and the correlations across the corresponding partitions. This suggests possible applications to less symmetric black-hole spacetimes, multi-horizon geometries, dynamical horizons, and configurations involving several competing island-like regions. Such generalizations could clarify how the weighting is modified by nontrivial topology, matter backreaction, or multiple entangling regions. The dependence of the weight on the cutoff surface also shows that its location and the associated definition of the subsystem boundary must be specified carefully in more general settings.

Higher-order replica contributions provide another important direction for future work. On the basis of the present structure, we expect the statistical weight to retain its functional form, while higher replicas modify the entropy through additional replica-dependent coefficients and corrections to the semiclassical action. Establishing this expectation requires an explicit evaluation of the higher-replica gravitational saddles and their associated fluctuation determinants. Such an analysis would determine the range of validity of the leading-order weighted expression and would also test whether the smooth crossover found here persists beyond the lowest replica approximation.

Our derivation was performed in Euclidean signature, where the relevant saddle-point expansion and convergence properties are more tractable. A Lorentzian extension is technically more involved because it requires a careful treatment of real-time replica geometries, causal structure, complex saddles, and the appropriate contour in the gravitational path integral. Nevertheless, the Lorentzian formulation is essential for describing the dynamical evolution of correlations during evaporation and for connecting the weighted prescription more directly to real-time information transfer. Extending the present framework to Lorentzian quantum gravity may therefore reveal a richer relation between the statistical competition of surfaces, entanglement-wedge reconstruction, and the dynamical Page curve.

Finally, the present derivation employs the on-shell action and assumes that fluctuations around the relevant semiclassical configurations are sufficiently suppressed. The formula is consequently most reliable in the thermodynamic regime, where the number of Hawking quanta is large and relative fluctuations are small. A more complete treatment should incorporate off-shell fluctuations, one-loop determinants, and the regime of a small number of Hawking particles. These extensions, together with higher-replica and Lorentzian analyses, will be necessary to determine the full domain of applicability of the weighted framework. Despite these limitations, our results indicate that retaining the statistical competition between candidate surfaces can provide a useful conceptual and technical extension of the standard QES prescription, while yielding a QES-like single-surface limit in the sharply peaked semiclassical regime.

\section*{Acknowledgments}
The authors would like to express their appreciation to Edward Witten for helpful discussions, for reading the manuscript, and for his valuable comments. They also thank Mohammad Hossein Zarei for his suggestion.
%%%%%%%%%%%%%%%%%%%%%%%%%%%%%%%%%%%%%%%%%%%%%%%%%%%%%%%%%%%%%%%%%%

\providecommand{\href}[2]{#2}\begingroup\raggedright\endgroup

\begin{thebibliography}{10}
	
	\bibitem{hawking1976a}
	S.W.~Hawking, \emph{Breakdown of predictability in gravitational collapse},
	\href{https://doi.org/10.1103/PhysRevD.14.2460}{\emph{Phys. Rev. D}
		{\bfseries 14} (1976) 2460}.
	
	\bibitem{ryu2006b}
	S.~Ryu and T.~Takayanagi, \emph{Holographic Derivation of Entanglement
				Entropy from AdS/CFT},
	\href{https://doi.org/10.1103/PhysRevLett.96.181602}{\emph{Phys. Rev. Lett.}
		{\bfseries 96} (2006) 181602}
	[\href{https://arxiv.org/abs/hep-th/0603001}{{\ttfamily hep-th/0603001}}].
	
	\bibitem{fursaev2006}
	D.V.~Fursaev, \emph{Proof of the Holographic Formula for Entanglement
				Entropy}, \href{https://doi.org/10.1088/1126-6708/2006/09/018}{\emph{J.
			High Energy Phys.} {\bfseries 2006} (2006) 018}
	[\href{https://arxiv.org/abs/hep-th/0606184}{{\ttfamily hep-th/0606184}}].
	
	\bibitem{nishioka2009b}
	T.~Nishioka, S.~Ryu and T.~Takayanagi, \emph{Holographic Entanglement
				Entropy: An Overview},
	\href{https://doi.org/10.1088/1751-8113/42/50/504008}{\emph{J. Phys. A: Math.
			Theor.} {\bfseries 42} (2009) 504008}
	[\href{https://arxiv.org/abs/0905.0932}{{\ttfamily 0905.0932}}].
	
	\bibitem{hubeny2007b}
	V.E.~Hubeny, M.~Rangamani and T.~Takayanagi, \emph{A Covariant Holographic
				Entanglement Entropy Proposal},
	\href{https://doi.org/10.1088/1126-6708/2007/07/062}{\emph{J. High Energy
			Phys.} {\bfseries 2007} (2007) 062}
	[\href{https://arxiv.org/abs/0705.0016}{{\ttfamily 0705.0016}}].
	
	\bibitem{faulkner2013b}
	T.~Faulkner, A.~Lewkowycz and J.~Maldacena, \emph{Quantum corrections to
		holographic entanglement entropy},
	\href{https://doi.org/10.1007/JHEP11(2013)074}{\emph{J. High Energ. Phys.}
		{\bfseries 2013} (2013) 74}
	[\href{https://arxiv.org/abs/1307.2892}{{\ttfamily 1307.2892}}].
	
	\bibitem{engelhardt2015b}
	N.~Engelhardt and A.C.~Wall, \emph{Quantum Extremal Surfaces: Holographic
				Entanglement Entropy beyond the Classical Regime},
	\href{https://doi.org/10.1007/JHEP01(2015)073}{\emph{J. High Energ. Phys.}
		{\bfseries 2015} (2015) 73}
	[\href{https://arxiv.org/abs/1408.3203}{{\ttfamily 1408.3203}}].
	
	\bibitem{almheiri2021b}
	A.~Almheiri, T.~Hartman, J.~Maldacena, E.~Shaghoulian and A.~Tajdini, \emph{The
		entropy of Hawking radiation},
	\href{https://doi.org/10.1103/RevModPhys.93.035002}{\emph{Rev. Mod. Phys.}
		{\bfseries 93} (2021) 035002}
	[\href{https://arxiv.org/abs/2006.06872}{{\ttfamily 2006.06872}}].
	
	\bibitem{page1983b}
	D.N.~Page, \emph{Comment on ``Entropy Evaporated by a Black Hole''},
	\href{https://doi.org/10.1103/PhysRevLett.50.1013}{\emph{Phys. Rev. Lett.}
		{\bfseries 50} (1983) 1013}.
	
	\bibitem{page1993a}
	D.N.~Page, \emph{Average Entropy of a Subsystem},
	\href{https://doi.org/10.1103/PhysRevLett.71.1291}{\emph{Phys. Rev. Lett.}
		{\bfseries 71} (1993) 1291}
	[\href{https://arxiv.org/abs/gr-qc/9305007}{{\ttfamily gr-qc/9305007}}].
	
	\bibitem{page2013a}
	D.N.~Page, \emph{Time Dependence of Hawking Radiation Entropy},
	\href{https://doi.org/10.1088/1475-7516/2013/09/028}{\emph{J. Cosmol.
			Astropart. Phys.} {\bfseries 2013} (2013) 028}
	[\href{https://arxiv.org/abs/1301.4995}{{\ttfamily 1301.4995}}].
	
	\bibitem{almheiri2019}
	A.~Almheiri, N.~Engelhardt, D.~Marolf and H.~Maxfield, \emph{{The entropy of
			bulk quantum fields and the entanglement wedge of an evaporating black
			hole}}, \href{https://doi.org/10.1007/JHEP12(2019)063}{\emph{J. High Energy
			Phys.} {\bfseries 2019} (2019) 63}
	[\href{https://arxiv.org/abs/1905.08762}{{\ttfamily 1905.08762}}].
	
	\bibitem{penington2020}
	G.~Penington, \emph{{Entanglement wedge reconstruction and the information
			paradox}}, \href{https://doi.org/10.1007/JHEP09(2020)002}{\emph{J. High
			Energy Phys.} {\bfseries 2020} (2020) 2}
	[\href{https://arxiv.org/abs/1905.08255}{{\ttfamily 1905.08255}}].
	
	\bibitem{Marolf2020}
	D.~Marolf, S.~Wang and Z.~Wang, \emph{{Probing phase transitions of holographic
			entanglement entropy with fixed area states}},
	\href{https://doi.org/10.1007/JHEP12(2020)084}{\emph{J. High Energy Phys.}
		{\bfseries 2020} (2020) } [\href{https://arxiv.org/abs/2006.10089}{{\ttfamily
			2006.10089}}].
	
	\bibitem{Dong2020}
	X.~Dong and H.~Wang, \emph{{Enhanced corrections near holographic entanglement
			transitions: a chaotic case study}},
	\href{https://doi.org/10.1007/JHEP11(2020)007}{\emph{J. High Energy Phys.}
		{\bfseries 2020} (2020) } [\href{https://arxiv.org/abs/2006.10051}{{\ttfamily
			2006.10051}}].
	
	\bibitem{Almheiri2020a}
	A.~Almheiri, R.~Mahajan, J.~Maldacena and Y.~Zhao, \emph{{The Page curve of
			Hawking radiation from semiclassical geometry}},
	\href{https://doi.org/10.1007/JHEP03(2020)149}{\emph{J. High Energy Phys.}
		{\bfseries 2020} (2020) 149}
	[\href{https://arxiv.org/abs/1908.10996}{{\ttfamily 1908.10996}}].
	
	\bibitem{Kudler-Flam2022}
	J.~Kudler-Flam and Y.~Kusuki, \emph{{On Quantum Information Before the Page
			Time}},  \href{https://arxiv.org/abs/2212.06839}{{\ttfamily 2212.06839}}.
	
	\bibitem{harlow2016}
	D.~Harlow, \emph{Jerusalem Lectures on Black Holes and Quantum
				Information},
	\href{https://doi.org/10.1103/RevModPhys.88.015002}{\emph{Rev. Mod. Phys.}
		{\bfseries 88} (2016) 015002}
	[\href{https://arxiv.org/abs/1409.1231}{{\ttfamily 1409.1231}}].
	
	\bibitem{dong2016}
	X.~Dong, \emph{The gravity dual of R\'enyi entropy},
	\href{https://doi.org/10.1038/ncomms12472}{\emph{Nat. Commun.}
		{\bfseries 7} (2016) 12472}
	[\href{https://arxiv.org/abs/1601.06788}{{\ttfamily 1601.06788}}].
	
	\bibitem{witten2025}
	E.~Witten, \emph{Introduction to Black Hole Thermodynamics}  [\href{https://arxiv.org/abs/2412.16795}{{\ttfamily 2412.16795}}].
	
	\bibitem{headrick2010}
	M.~Headrick, \emph{Entanglement Renyi entropies in holographic theories},
	\href{https://doi.org/10.1103/PhysRevD.82.126010}{\emph{Phys. Rev. D}
		{\bfseries 82} (2010) 126010}
	[\href{https://arxiv.org/abs/1006.0047}{{\ttfamily 1006.0047}}].
	
	\bibitem{almheiri2020b}
	A.~Almheiri, T.~Hartman, J.~Maldacena, E.~Shaghoulian and A.~Tajdini,
	\emph{Replica Wormholes and the Entropy of Hawking Radiation},
	\href{https://doi.org/10.1007/JHEP05(2020)013}{\emph{J. High Energ. Phys.}
		{\bfseries 2020} (2020) 13}
	[\href{https://arxiv.org/abs/1911.12333}{{\ttfamily 1911.12333}}].
	
	\bibitem{penington2022b}
	G.~Penington, S.H.~Shenker, D.~Stanford and Z.~Yang, \emph{Replica wormholes
		and the black hole interior},
	\href{https://doi.org/10.1007/JHEP03(2022)205}{\emph{J. High Energ. Phys.}
		{\bfseries 2022} (2022) 205}
	[\href{https://arxiv.org/abs/1911.11977}{{\ttfamily 1911.11977}}].
	
	\bibitem{mertens2023}
	T.G.~Mertens and G.J.~Turiaci, \emph{Solvable Models of Quantum Black
				Holes: A Review on Jackiw-Teitelboim Gravity},
	\href{https://doi.org/10.1007/s41114-023-00046-1}{\emph{Living Rev Relativ}
		{\bfseries 26} (2023) 4} [\href{https://arxiv.org/abs/2210.10846}{{\ttfamily
			2210.10846}}].
		
			\bibitem{Marolf2021}
		D.~Marolf and H.~Maxfield, \emph{{Observations of Hawking radiation: the Page
				curve and baby universes}},
		\href{https://doi.org/10.1007/JHEP04(2021)272}{\emph{J. High Energy Phys.}
			{\bfseries 2021} (2021) } [\href{https://arxiv.org/abs/2010.06602}{{\ttfamily
				2010.06602}}].
		
		\bibitem{Marolf2021a}
		D.~Marolf and H.~Maxfield, \emph{{The page curve and baby universes}},
		\href{https://doi.org/10.1142/S021827182142027X}{\emph{Int. J. Mod. Phys. D}
			{\bfseries 30} (2021) } [\href{https://arxiv.org/abs/2105.12211}{{\ttfamily
				2105.12211}}].
		
		\bibitem{Geng2021}
		H.~Geng, Y.~Nomura and H.Y.~Sun, \emph{{Information paradox and its resolution
				in de Sitter holography}},
		\href{https://doi.org/10.1103/PhysRevD.103.126004}{\emph{Phys. Rev. D}
			{\bfseries 103} (2021) 126004}
		[\href{https://arxiv.org/abs/2103.07477}{{\ttfamily 2103.07477}}].
		
		\bibitem{Arefeva2021}
		I.~Aref'eva and I.~Volovich, \emph{{A Note on Islands in Schwarzschild Black
				Holes}},  \href{https://arxiv.org/abs/2110.04233}{{\ttfamily 2110.04233}}.
		
		\bibitem{Hashimoto2020}
		K.~Hashimoto, N.~Iizuka and Y.~Matsuo, \emph{{Islands in Schwarzschild black
				holes}}, \href{https://doi.org/10.1007/JHEP06(2020)085}{\emph{J. High Energy
				Phys.} {\bfseries 2020} (2020) }
		[\href{https://arxiv.org/abs/2004.05863}{{\ttfamily 2004.05863}}].
		
		\bibitem{Wang2021}
		X.~Wang, R.~Li and J.~Wang, \emph{{Page curves for a family of exactly solvable
				evaporating black holes}},
		\href{https://doi.org/10.1103/PhysRevD.103.126026}{\emph{Phys. Rev. D}
			{\bfseries 103} (2021) 126026}.
		
		\bibitem{Gautason2020}
		F.F.~Gautason, L.~Schneiderbauer, W.~Sybesma and L.~Thorlacius, \emph{{Page
				curve for an evaporating black hole}},
		\href{https://doi.org/10.1007/JHEP05(2020)091}{\emph{J. High Energ. Phys.}
			{\bfseries 2020} (2020) } [\href{https://arxiv.org/abs/2004.00598}{{\ttfamily
				2004.00598}}].
			
				\bibitem{hartman2020a}
			T.~Hartman, E.~Shaghoulian and A.~Strominger, \emph{Islands in Asymptotically Flat 2D Gravity}, \href{https://doi.org/10.1007/JHEP07(2020)022}{\emph{J.
					High Energ. Phys.} {\bfseries 2020} (2020) 22}
			[\href{https://arxiv.org/abs/2004.13857}{{\ttfamily 2004.13857}}].
	
	\bibitem{lewkowycz2013b}
	A.~Lewkowycz and J.~Maldacena, \emph{Generalized gravitational entropy},
	\href{https://doi.org/10.1007/JHEP08(2013)090}{\emph{J. High Energ. Phys.}
		{\bfseries 2013} (2013) 90}
	[\href{https://arxiv.org/abs/1304.4926}{{\ttfamily 1304.4926}}].
	
	\bibitem{Akers2021}
	C.~Akers and G.~Penington, \emph{{Leading order corrections to the quantum
			extremal surface prescription}},
	\href{https://doi.org/10.1007/JHEP04(2021)062}{\emph{J. High Energy Phys.}
		{\bfseries 2021} (2021) } [\href{https://arxiv.org/abs/2008.03319}{{\ttfamily
			2008.03319}}].
	
	\bibitem{blommaert2023}
	A.~Blommaert, J.~Kruthoff and S.~Yao, \emph{The power of Lorentzian
		wormholes}, \href{https://doi.org/10.1007/JHEP10(2023)005}{\emph{J. High
			Energ. Phys.} {\bfseries 2023} (2023) 5}.
	
	\bibitem{engelhardt2023}
	N.~Engelhardt and H.~Liu, \emph{Algebraic ER=EPR and Complexity
				Transfer}, \href{https://doi.org/10.1007/JHEP07(2024)013}{\emph{J. High Energ. Phys.}
			{\bfseries 2024} (2024) 13}
		[\href{https://arxiv.org/abs/2311.04281}{{\ttfamily 2311.04281}}].
	
	\bibitem{engelhardt2022}
	N.~Engelhardt and {\AA}.~Folkestad, \emph{Canonical Purification of
		Evaporating Black Holes},
	\href{https://doi.org/10.1103/PhysRevD.105.086010}{\emph{Phys. Rev. D}
		{\bfseries 105} (2022) 086010}
	[\href{https://arxiv.org/abs/2201.08395}{{\ttfamily 2201.08395}}].
	
	\bibitem{maldacena2013}
	J.~Maldacena and L.~Susskind, \emph{Cool horizons for entangled black holes},
	\href{https://doi.org/10.1002/prop.201300020}{\emph{Fortschritte der Physik}
		{\bfseries 61} (2013) 781} [\href{https://arxiv.org/abs/1306.0533}{{\ttfamily
			1306.0533}}].
	
	\bibitem{antonini2025}
	S.~Antonini, C.-H.~Chen, H.~Maxfield and G.~Penington, \emph{An apologia for
		islands} [\href{https://arxiv.org/abs/2506.04311}{{\ttfamily
			2506.04311}}].
	
	\bibitem{raamsdonk2010a}
	M.V.~Raamsdonk, \emph{Building up spacetime with quantum entanglement},
	\href{https://doi.org/10.1007/s10714-010-1034-0
		10.1142/S0218271810018529}{\emph{Gen Relativ Gravit} {\bfseries 42} (2010)
		2323} [\href{https://arxiv.org/abs/1005.3035}{{\ttfamily 1005.3035}}].
	
	\bibitem{raamsdonk2010}
	M.V.~Raamsdonk, \emph{Comments on quantum gravity and entanglement}  [\href{https://arxiv.org/abs/0907.2939}{{\ttfamily 0907.2939}}].
	
	\bibitem{wu2022}
	Shu-Min Wu and Hao-Sheng Zeng \emph{Genuine tripartite nonlocality and entanglement in curved spacetime},
	\href{https://doi.org/10.1140/epjc/s10052-021-09954-4}{\emph{Eur. Phys. J. C}
		{\bfseries 82} (2022) 4}
	[\href{https://arxiv.org/abs/2201.02333}{{\ttfamily 2201.02333}}].
	
	\bibitem{wu2023}
	Shu-Min Wu, Xiao-Wei Teng, Jin-Xuan Li, Si-Han Li, Tong-Hua Liu and Jie-Ci Wang \emph{Genuinely accessible and inaccessible entanglement in Schwarzschild black hole},
	\href{https://doi.org/10.1016/j.physletb.2023.138334}{\emph{Phys. Lett. B}
		{\bfseries 848} (2024) 138334}
	[\href{https://arxiv.org/abs/2311.12362}{{\ttfamily 2311.12362}}].
	
	\bibitem{wu2024}
	Shu-Min Wu and Xiao-Wei Teng and Hao-Yu Wu and Jin-Xuan Li and Xiao-Li Huang and Rui Bao \emph{Conserved mutual information for discrete and continuous variables in dilaton black hole},
	\href{https://doi.org/10.1016/j.cjph.2024.09.039}{\emph{Chin. J. Phys.}
		{\bfseries 92} (2024) 755}.
	
\end{thebibliography}
\end{document}